\documentclass[aps,footinbib,prd,superscriptaddress,reprint]{revtex4-1}

\usepackage{graphicx}
\usepackage{amssymb}

\usepackage{amsmath}
\usepackage{slashed}
\usepackage{lipsum}
\usepackage[dvipsnames]{xcolor}
\usepackage{xcolor}   
\usepackage{xspace} 
\usepackage{bm}
\usepackage{xfrac}
\usepackage{lineno}
\usepackage[normalem]{ulem}
\usepackage{comment}
\usepackage{tikz}
\usetikzlibrary{arrows, backgrounds, calc, decorations.pathreplacing, decorations.markings, fadings, quotes, positioning, shadings, snakes, intersections, angles, patterns}
\usepackage{footmisc}

\usepackage[T1]{fontenc}      
\usepackage{ae} 
\usepackage{grffile}

\newcommand{\im}{\ensuremath{\mathrm{i}}}

\usepackage[colorlinks=true]{hyperref}
\hypersetup{
    bookmarks=true,         
    unicode=false,          
    pdftoolbar=true,        
    pdfmenubar=true,        
    pdffitwindow=false,     
    pdfstartview={FitH},    
    pdfnewwindow=true,      
    colorlinks=true,        
    linkcolor=blue,         
    citecolor=red,          
    filecolor=magenta,      
    urlcolor=blue           
}

\newcommand{\p}	{\partial}

\newcommand{\C}	{\mathbb{C}}

\newcommand{\cA}{\mathcal{A}}

\newcommand{\cN}{\mathcal{N}}
\newcommand{\cO}{\mathcal{O}}

\newcommand{\cU}{\mathcal{U}}
\newcommand{\cV}{\mathcal{V}}

\newcommand{\sfH}{\mathsf{H}}

\newcommand{\bfk}   {\mathbf{k}}
\newcommand{\bfq}   {\mathbf{q}}
\newcommand{\bfp}   {\mathbf{p}}

\DeclareMathOperator{\tr}{\tr}

\newcommand\x{\vec x}

\newcommand\A{\mathbf{A}}

\newcommand{\bra}[1]	{\langle{#1}\vert}
\newcommand{\ket}[1]	{\vert{#1}\rangle}
\newcommand{\braket}[2]	{\langle{#1}\vert{#2}\rangle}
\newcommand{\corr}[1]{\left\langle{#1}\right\rangle}
\newcommand{\ketbra}[2]	{\ket{#1}\bra{#2}}

\newcommand{\po}{\hat{r}}
\newcommand{\vo}{\hat{v}}
\newcommand{\bz}{\bar{z}}
\newcommand{\lcb}{\xi}
\newcommand{\WP}{\mathsf{W}}
\newcommand{\WK}{\mathsf{K}}
\newcommand{\WR}{\mathsf{R}}
\newcommand{\WH}{\mathsf{H}}

\newcommand{\m}{\mathsf{m}}

\renewcommand{\tr}	{\mathrm{tr}}
\newcommand{\Mat}   {\mathsf{Mat}}

\newcommand\id		{\mathbf{1}}

\newcommand{\LB}    {l_B}

\newcommand{\ZB}{\mathsf{w}}

\newcommand{\tepsilon}{\tilde{\epsilon}}
\newcommand{\tcA}   {\tilde{\cA}}
\newcommand{\tOmega}{\tilde{\Omega}}
\newcommand{\tu}    {\tilde{u}}

\newcommand{\tg}    {\tilde{g}}
\newcommand{\tD}    {\tilde{D}}

\newcommand{\tU}    {\tilde{U}}
\newcommand{\tV}    {\tilde{V}}
\newcommand{\tcU}{\tilde{\cU}}
\newcommand{\tcV}{\tilde{\cV}}
\newcommand{\tZB}{\tilde{\ZB}}
\newcommand{\tH}{\tilde{H}}
\newcommand{\pertket}{\widetilde{\ket{\bfk\alpha}}}

\newcommand{\e}{\mathrm{e}}
\renewcommand{\d}{\mathrm{d}}

\newcommand{\FBZ}{\text{FBZ}}
\renewcommand{\th}{\mathrm{th}}
\renewcommand{\Re}{\text{Re}}
\renewcommand{\Im}{\text{Im}}

\newcommand\quotes[1]  {``{#1}"}

\newcommand\Secref[1]	{Section~\ref{#1}\xspace}

\newcommand\Figref[1]	{Figure~\ref{#1}\xspace}

\newcommand\secref[1]	{section~\ref{#1}\xspace}

\newcommand\figref[1]	{figure~\ref{#1}\xspace}
\newcommand\appref[1] {appendix~\ref{#1}\xspace}

\begin{document}

\title{Interplay of Band Geometry and Topology in Ideal Chern Insulators in Presence of External Electromagnetic Fields}



\affiliation{Institute for Theoretical Physics,
Julius-Maximilians-Universit\"at W\"urzburg, 97074 W\"urzburg, Germany}

\affiliation{W\"urzburg-Dresden Cluster of Excellence ct.qmat}

\affiliation{School of Theoretical Physics, Dublin Institute for Advanced Studies, 10 Burlington Road, Dublin 4, Ireland}

\author{Christian~Northe}

\email{christian.northe@physik.uni-wuerzburg.de}

\affiliation{Institute for Theoretical Physics, Julius-Maximilians-Universit\"at W\"urzburg, 97074 W\"urzburg, Germany}

\affiliation{W\"urzburg-Dresden Cluster of Excellence ct.qmat}

\author{Giandomenico~Palumbo}

\affiliation{School of Theoretical Physics, Dublin Institute for Advanced Studies, 10 Burlington Road, Dublin 4, Ireland}

\author{Jonathan~Sturm}

\affiliation{Institute for Theoretical Physics,
Julius-Maximilians-Universit\"at W\"urzburg, 97074 W\"urzburg, Germany}

\author{Christian~Tutschku}

\affiliation{Fraunhofer-Institut für Arbeitswirtschaft und Organisation , 70569 Stuttgart, Germany}

\author{Ewelina.~M.~Hankiewicz} 

\email{ewelina.hankiewicz@physik.uni-wuerzburg.de}

\affiliation{Institute for Theoretical Physics,
Julius-Maximilians-Universit\"at W\"urzburg, 97074 W\"urzburg, Germany}

\affiliation{W\"urzburg-Dresden Cluster of Excellence ct.qmat}

\begin{abstract}
Ideal Chern insulating phases arise in two-dimensional systems with broken time-reversal symmetry. They are characterized by having nearly-flat bands, and a uniform quantum geometry -- which combines the Berry curvature and quantum metric -- and by being incompressible. In this work, we analyze the role of the quantum geometry in ideal Chern insulators aiming to describe transport in presence of external out-of-plane magnetic and electric fields. We firstly show that in the absence of external perturbations, novel Berry connections appear in ideal Chern insulating phases. Secondly, we provide a detailed analysis of the deformation of the quantum geometry once weak out-of-plane magnetic fields are switched on. 
The perturbed Berry curvature and quantum metric provide an effective quantum geometry, which is analyzed in the insulating regime and provides an application of our novel connections. The conditions under which the Girvin-MacDonald-Platzman algebra is realized in this situation are discussed. Furthermore, an investigation  of electrical transport due to the new effective quantum geometry is presented once an electric field is added. Restricting to the case of two bands in the metallic regime the quantum metric appears as measurable quantum mechanical correction in the Hall response. Our findings can be applied, for instance, to rhombohedral trilayer graphene at low energies. 

\end{abstract}




\maketitle


\section{Introduction}

Since the seminal work by Haldane \cite{PhysRevLett.61.2015}, Chern insulators have been recognized as prime examples of topological phases of matter.
The reason is their natural relation to the quantum Hall effect (QHE), in which a strong external magnetic field, such that the cyclotron orbits are much smaller than the system size, gives rise to Landau levels. Here, the quantum Hall states (QHs) are characterized by Chern numbers that induce topologically protected chiral edge states \cite{PrangeGirvin}.
Similarly, Chern insulators are characterized by Chern numbers in the bulk and gapless chiral edge modes on the boundary, due to the bulk-boundary correspondence \cite{WenBook}. In contrast to the QHs however, their topological bulk states are not induced by any external magnetic field. Rather, the Berry curvature of the Bloch bands substitutes the magnetic field and the conventional flat Landau levels (LLs) are replaced by dispersive Bloch bands. Due to the absence of LLs, the density operators in the Chern insulators do not, in general, satisfy the Girvin-MacDonald-Platzman (GMP) algebra, which is one of the main pillars of the QHs and explains the incompressibility of the Hall fluids and the existence of the magnetorotor mode \cite{PhysRevB.33.2481} in the fractional QHE (FQHE). 


One modern approach to quantify the properties of topological phases, and Chern insulators in particular, is via the quantum geometric tensor \cite{Resta_2011}. Its imaginary part is the well established Berry curvature characterizing the topological phase of the band, while its real part is the quantum metric. The study of the quantum metric, and thus also the quantum geometric tensor, has attracted attention on many fronts in the recent past including superconducting systems \cite{PhysRevB.103.014516, PhysRevLett.124.197002, herzogarbeitman, Peotta}, quantum phase transitions \cite{PhysRevLett.99.100603, Panahiyan}, magnetic signatures \cite{PhysRevB.91.214405, Scaffidi, bauer2021fractional, rhim}, quantum topology and geometry \cite{koziiIntrinsic, Ozawa_2021, Mera_2021, PalumboCigar, ahn2021riemannian}, flat band systems \cite{Mitscherling}, in non-adiabatic evolution \cite{Bleu}, as marker distinguishing insulators from metals \cite{Marrazzo}, non-hermitian systems \cite{Zhu, LeeNonHermitian}, in twisted bilayer graphene \cite{Xie, Rossi, Julku, Wu, AhnSuperconductivity, Ledwith, Toermae, Hu} and in dimensions higher than two \cite{NeupertNCG, PalumboTensor, PalumboFloquet, Ansgar, Pozo, Lin, meraZhang, kemp}. Theoretical measurements have been proposed and measurements of the quantum geometric tensor have been performed in \cite{Yu_2019, OzawaGoldman, BleuMeasurement, Gersdorff, Tan, TanSuperconducting, chensynthetic, gianfrate}. On top of this, we provide in this paper further measurable consequences of the quantum geometry layed bare by magnetic fields.

As shown in \cite{Parameswaran_2012, Roy_2014}, special types of Chern insulators, so-called \textit{ideal isotropic Chern bands}, manage to mimick LLs. Such bands are characterized mainly by three features:
\begin{itemize}
  \item [$(a)$] They are nearly-flat, isotropic and well separated from other bands.
  \item [$(b)$] Across the first Brillouin zone ($\FBZ$) these bands have a nearly-uniform quantum geometric tensor.
  \item [$(c)$] They are so-called isotropic droplets, a notion to be made mathematically precise below. Physically, these droplets are equivalent to incompressible fluids.
\end{itemize}
In the following we always imply the specifier \quotes{nearly} when discussing flat bands or uniform quantum geometry.
Equipped with these three properties ideal Chern bands resemble formally a LL where the Berry curvature takes on the role of the magnetic field and the quantum metric the role of the Galilean metric employed by Haldane \cite{Haldane_2011} to argue in favor of possible \quotes{gravitational} degrees of freedom in the FQHE. Such bands even realize a GMP algebra to all orders in momentum \cite{Roy_2014}. Therefore ideal Chern bands are optimal candidates to emulate the FQHE in lattice systems \cite{RegnaultBernevig, Neupert_2015, ThomaleDuality} and in absence of strong magnetic fields. Even though lattice systems never fulfill all three requirements of an ideal isotropic Chern band simultaneously and exactly \cite{Varjas}, there are examples which feature these three properties to a remarkable degree \cite{Lee_2017}, so that all benefits of LLs are available in good approximation. 

The purpose of this paper is to investigate the properties of ideal isotropic Chern bands in the presence of weak magnetic fields and, possibly inhomogeneous, electric fields. We proceed in three major steps. Firstly, before implementing electromagnetic fields, we revisit the property of uniformity in ideal Chern bands (item $(b)$ above) by demonstrating the existence of novel emergent Berry connections associated with Bloch bands, which allow to quantify the uniformity of their corresponding quantum geometry's uniformity. Moreover, these connections give rise to quantized Chern numbers should the band indeed have uniform quantum geometry. 

In the second step, we analyze ideal Chern insulators in presence of weak external out-of-plane magnetic fields. In the following we always imply the specifier \quotes{out-of-plane} when discussing magnetic fields. The only vital ingredient in this analysis is in fact the flatness of the band (property $(a)$ above). The missing dispersion in the band allows to perform perturbation theory for small magnetic perturbations. This unveils the existence of an effective quantum geometry, i.e. the deformation of the Berry curvature and quantum metric due to a weak magnetic field. Importantly, the corrections are fully of quantum mechanical origin and thus not accessible in conventional semiclassical analyses. We proceed to study the consequences of this new quantum geometry in the insulating regime. Our results demonstrate that the magnetic field does not destroy the topology, even after including quantum mechanical corrections. This is similar to \cite{BoettcherTutschku,TutschkuBoettcher, Tutschku} where magnetic field does not foil the quantum anomalous Hall effect. Moving on we study the effect of the magnetic field on the three properties of ideal isotropic Chern bands. This provides an application of the emergent Berry connections discussed prior, since the magnetic field affects the uniformity of the quantum geometry. Moreover, we show to which extend and under what conditions the GMP algebra is still realized, which is of importance in future studies implementing interactions.

In a third and final step, we analyze electrical transport due to the effective quantum geometry in the metallic regime by implementing an inhomogeneous electric field \cite{Lapa_2019}. We present a semiclassical analysis based on our new effective quantum geometry and derive the electrical current. It contains two contributions, the Hall current and the geometric current. Both pick up measurable corrections linear in the magnetic field as a consequence of the effective quantum geometry. We exemplify our formalism using a two-band model.

Our work paves the way for a deeper understanding of ideal Chern insulators in presence of external fields and presents the starting point to study the interplay between the quantum metric and external electromagnetic fields in the fractional insulating phases. Moreover, we emphasize that our perturbation theory relies only on the presence of a flat band, so that it may be applied to a larger class of models than ideal bands. One candidate is rhombohedral trilayer graphene at low energies \cite{Mitscherling}, which harbors two flat bands.

This paper is organized as follows. In \secref{sec: Recap} we recapulate briefly the notion quantum geometry and that of ideal isotropic bands. Thereafter, in \secref{sec: NewConnections}, we introduce the emergent Berry connections quantifying the uniformity of the band's quantum geometry. \Secref{sec: weakB} provides the core of this article, presenting the perturbation theory which weaves the weak magnetic field into the quantum geometry. Subsequently, we analyze this quantum geometry in the insulating regime in \secref{sec: IdealInsulatorsB} and electrical transport in the metallic regime in \secref{sec: Metals}. A two-band example is provided in \secref{sec: Example} before discussing our results in \secref{sec: conclusions}. Detailed appendices provide involved calculations.

\section{Quantum geometry in Chern bands}\label{sec: Recap}
In this section, we briefly summarize the main features of ideal Chern insulators by focusing on their band geometry.
The corresponding tight-binding lattice models carry topologically non-trivial Bloch bands. 
Placing the chemical potential in the gap and invoking Bloch's theorem, the single particle Hamiltonian of an $\cN$ band insulator is described by
\begin{equation}\label{BlochHamiltonian}
 H=\sum_\bfk\sum_{ab}\ket{\bfk,a}\,h_{ab}(\bfk)\,\bra{\bfk,b}, 
\end{equation}
Throughout, momenta range within the first Brillouin zone ($\FBZ$), $\{a, b\}=1,\dots,\cN$ label orbitals within the $n^\th$ unit cell. For each band $\alpha=1,\dots, \cN$ the Bloch Hamiltonian $h_{ab}(\bfk)\in\Mat(\cN, \C)$ satisfies the eigenvalue problem $\sum_{b}h_{ab}(\bfk)\,u^\alpha_b(\bfk)=\epsilon_\alpha(\bfk)u^\alpha_a(\bfk)$
%
%
where $\epsilon_\alpha$ is the energy of the $\alpha^\th$ band and $u^\alpha_a$ is the corresponding Bloch state with normalizations $\sum_a\,u^{\alpha *}_a(\bfk)u^\beta_a(\bfk)=\delta^{\alpha\beta}$ and $\sum_\alpha\,u^{\alpha *}_a(\bfk)u^\alpha_b(\bfk)=\delta_{ab}$.
%
%
The Bloch wave functions give rise to $\cN$ eigenstates 
\begin{equation}\label{BlochEigenstates}
 \ket{\bfk,\alpha}:=\sum_a\,u^\alpha_a(\bfk)\,\ket{\bfk,a}, 
\end{equation}
with orthogonality relation $\braket{\bfk,\alpha}{\bfq,\beta}=\delta(\bfk-\bfq)\delta^{\alpha\beta}$. They spectrally decompose the Hamiltonian, $ H_0
 =
 \sum_{\bfk}\sum_{\alpha=1}^\cN \epsilon_\alpha(\bfk)\ket{\bfk,\alpha}\bra{\bfk,\alpha}$.
The matrix elements of the position operator in this basis are
\begin{equation}\label{POband}
 \bra{\bfq \beta}\po_j\ket{\bfk\alpha}
 =
 \im\delta^{\beta\alpha}\p_j(\delta(\bfq-\bfk))+\delta(\bfq-\bfk)\cA^{\beta\alpha}_j(\bfk)
\end{equation}
where $\p_j=\p/\p k^j$, $j=\{x,y\}$ and 
\begin{equation}\label{CrossGap}
 \cA^{\beta\alpha}_j(\bfk)=\im\sum_{b=1}^{\cN}\,u^{\beta *}_b(\bfk)\p_j\, u_b^\alpha(\bfk), 
 \quad (\cA^{\beta\alpha}_j)^*=\cA^{\alpha\beta}_j\,.
\end{equation}
is the cross-gap function. When $\beta\neq\alpha$ the cross-gap function is a gauge invariant. In contrast, $\cA^{\alpha\alpha}\equiv\cA^\alpha$ is the Berry connection of the band. Since we work with non-degenerate bands the gauge group is $U(1)$. Each band comes equipped with a quantum geometric tensor
\begin{subequations}\label{QG}
\begin{align}
 \chi_{ij}^\alpha(q)
 &=
 \sum_{ab}(\p_i u^{\alpha *}_a)Q^\alpha_{ab}(q)(\p_j u^\alpha_b)\label{QGstandard}\\
 &=
 \sum_{\beta\neq\alpha}\cA^{\alpha\beta}_i(q)\cA^{\beta\alpha}_j(q)\label{QGband}\\
 &=
 \sum_a(D_i u^\alpha_a)^*(D_j u^\alpha_a)\label{QGcovariant}
\end{align}
\end{subequations}
where $D_j=\p_j+\im\cA_j(\bfk)$ is the gauge covariant derivative and $Q^\alpha_{ab}(\bfk)=\delta_{ab}-P^\alpha_{ab}(\bfk)$ and $P^\alpha_{ab}(\bfk)=u^\alpha_a(\bfk)u^{\alpha *}_b(\bfk)$ are projectors onto Bloch functions. The first line is the standard phrasing of $\chi^\alpha$. The real and imaginary parts of the quantum geometry are the quantum metric and Berry curvature respectively 
\begin{align}\label{BerryCurvature}
%
g^\alpha_{ij}(\bfk)
&=
\Re\left(\chi^\alpha_{ij}(\bfk)\right)=\sum_a(\p_{(i} u^{\alpha *}_a)(\p_{j)} u^\alpha_a)-\cA^\alpha_{(i}\cA^\alpha_{j)}\notag\\
%
%
\Omega^\alpha_{ij}(\bfk) 
&=-2\Im\left(\chi^\alpha_{ij}(\bfk)\right)=2\p_{[i}\cA^\alpha_{j]}\,.
\end{align}
with (anti-)symmetrization prescriptions $2A_{(i}B_{j)}=A_i B_j+A_j B_i$ and $2A_{[i}B_{j]}=A_i B_j-A_j B_i$. The Chern number for the $\alpha^\th$ band results from integration,
\begin{equation}
 C_\alpha
 =
 \frac{1}{2\pi}\int_{\FBZ}\d^2k\,\Omega^\alpha(\bfk).
\end{equation}
where $\Omega^\alpha(\bfk)=\varepsilon^{ij}\Omega^\alpha_{ij}/2$ is the only non-trivial component of the Berry curvature and $\varepsilon^{ij}$ the Levi-Civita symbol. A filled band with Chern number $C_\alpha$ gives rise to a Hall conductance $\sigma_{\text{H}}=\frac{e^2}{h}C_\alpha$. Whenever $C_\alpha\neq0$ we speak of a Chern band.

\subsection{Ideal Isotropic Chern Bands}
\textit{Ideal isotropic Chern bands} are constructed to emulate Landau levels. A Chern band $\alpha$ achieves this when, for $\bfk\in \FBZ$, it possesses three properties
\begin{itemize}
 \item[$(a)$] \textit{It is an isolated flat band}\footnote{\label{foot: subtleltyFlatBands}We note that topologically non-trivial flat bands do not really exist in two dimensions, so that we should write $\p_j\epsilon_\alpha(\bfk)\ll|\epsilon_\beta-\epsilon_\alpha|$ rather than $\p_j\epsilon_\alpha(\bfk)=0$. This is the distinction between nearly-flat bands and truly flat bands.}
 \begin{subequations}
 \begin{align}
  \p_j\epsilon_\alpha(\bfk)&=0,\quad |\epsilon_\beta(\bfk)-\epsilon_\alpha|\gg1\label{isolatedFlat}%
\intertext{\item[$(b)$] \textit{It has uniform quantum geometry} }%
        \chi^\alpha_{ij}(\bfk)&\approx const, \label{uniformity}%
\intertext{\item[$(c)$] \textit{It is an isotropic droplet, i.e. it satisfies}}%
     \tr\, g^\alpha(\bfk)&=|\Omega^\alpha(\bfk)|\label{droplet}
 \end{align}
 \end{subequations}
\end{itemize}
The constraint \eqref{droplet} enforces $g^\alpha_{xx}=g^\alpha_{yy}=\Omega^\alpha/2$ and $g^\alpha_{xy}=g^\alpha_{yx}=0$, hence the epiphet \quotes{isotropic}. Note that $g^\alpha_{ij}$ is unrelated to the metric describing real space. Since the quantum metric quantifies how a much a state rotates when moving between points in the $\FBZ$, this isotropy is a quality of Hilbert space.

It has been shown \cite{Roy_2014} that a Chern band with the properties \eqref{isolatedFlat}-\eqref{droplet} admits the GMP algebra \cite{PhysRevB.33.2481},
\begin{equation}\label{GMP}
 [\rho_{\bfq}^{\alpha},\rho_{\bfk}^{\alpha}]
 =
 2\im\,\sin \left( \frac{q_i\varepsilon^{ij}k_j\,\Omega^\alpha}{2} \right) \, 
 \exp\bigl[ q^i  g^\alpha_{ij}  k^j\bigr]\,
 \rho_{\bfq+\bfk}^{\alpha}
\end{equation}
to all orders in momentum; if only properties $(a)$ and $(b)$ hold, the band still satisfies \eqref{GMP} at long wavelengths \cite{Parameswaran_2012}. Here $\rho_{\bfk}^{\alpha}=P_\alpha \rho_\bfk P_\alpha=P_\alpha \exp(\im k_j\po^j) P_\alpha$ is the projected density operator and $P_\alpha=\sum_\bfk\ket{\bfk\alpha}\bra{\bfk\alpha}$ the band projector. Throughout this text, we employ Einstein's summation convention on momentum indices $i,j$. The algebra \eqref{GMP} is one of the hallmarks of LLs giving way for the \quotes{identification} of the ideal Chern band with a Landau level. The GMP algebra represents a quantum deformation of area-preserving diffeomorphisms. Hence it relfects the incompressibility of the systems, when viewed as a fluid.

\section{Emergent Berry connections through uniform quantum geometry}
\label{sec: NewConnections}
Here, we introduce novel Berry connections for Chern insulators, which emerge from the uniformity of the Bloch band's quantum geometry. As we show below, their corresponding curvature Berry tensors become relevant to quantify the uniformity by giving rise to quantized Chern numbers. In this section we introduce the necessary formalism; an example follows in \secref{sec: Example}.

It is useful to introduce the complex zweibein \cite{Dobardi, ahn2021riemannian}
\begin{equation}\label{CovariantDerivative}
 \ZB_{ja}(\bfk) 
 :=
 \sum_bQ^\alpha_{ab} (\p_ju^\alpha_b)
 \equiv
 (D_ju^\alpha_a)
\end{equation}
which is the \quotes{square root} of the quantum geometric tensor \eqref{QGcovariant} for the $\alpha^\th$ band, 
\begin{equation}
 \chi^\alpha_{ij}=\sum_a\ZB_{ia}^*\ZB_{ja}\equiv \ZB^\dagger_i\cdot\ZB_j 
\end{equation}
and $g^\alpha_{ij}=\ZB^\dagger_{(i}\cdot\ZB_{j)}$ and $\Omega^\alpha=\im\varepsilon^{ij}\ZB^\dagger_{i}\cdot\ZB_{j}$, where the scalar product encodes the summation over the orbital label. 
The zweibein allows to define
\begin{equation}
 Z_{ij}{}^l=\im\ZB_i^\dagger\cdot\p^l\ZB_j.
\end{equation}
The lower two indices are to be thought of as separate from the upper, as becomes evident under gauge transformations $\ZB_j\to \e^{\im\phi}\ZB_j$ for which $Z_{ij}{}^l \to Z_{ij}{}^l-\chi_{ij}(\bfk)\,\p^l\phi$. When the quantum geometry is uniform, $\chi(\bfk)=\chi$, it can be absorbed into the gauge function $\phi$, prompting the definition of two new $U(1)$ connections,
\begin{subequations}\label{NovelConnections}
\begin{align}
 U^l&=\delta^{ij}Z_{ij}{}^l
 =
 i\ZB_j^\dagger\cdot\p^l\ZB^j\label{MetricConnection}\\
 V^\lambda&=\frac{1}{\im}\epsilon^{ij}Z_{ij}{}^l
 =
 \epsilon^{ij}\ZB_i^\dagger\cdot\p^l\ZB_j\label{CurvatureConnection}
\end{align}
\end{subequations}
Their transformation properties are $U^l\to U^l-\tr(g^\alpha)\,\p^l\phi$ and $V^l\to V^l+\Omega^\alpha\,\p^l\phi$. Hence, they measure individually the uniformity of $g^\alpha$ and $\Omega^\alpha$, respectively, while $Z_{ij}{}^l$ measures that of the full quantum geometry.
We stress that \eqref{NovelConnections} become $U(1)$ connections only in the case of uniform quantum geometry. As such they probe the uniformity of the band's quantum geometry. 

In analogy to the Berry curvature, we introduce exterior derivatives of \eqref{NovelConnections},
\begin{subequations}\label{NovelCurvatures}
\begin{align}
 \cU^{lk}=\p^l U^k-\p^k U^l,\label{CurvatureU}\\
 \cV^{lk}=\p^l V^k-\p^k V^l\label{CurvatureV}\, ,
\end{align}
\end{subequations}
If the quantum geometry is uniform, these tensors are gauge-invariant curvatures and their integration over the $\FBZ$ give rise to a quantized Chern numbers, since the $\FBZ$ is a torus.

\section{Flat bands in constant magnetic field}\label{sec: weakB}
Here, we study the geometric response of ideal Chern insulators in presence of a constant, uniform magnetic field $B$. Notice that we consider here a magnetic field not strong enough to push the Chern insulating system into the Hofstadter phase. The magnetic field $B$ does not need to have rational flux. In general, a magnetic perturbation shifts the momentum via the Peierls substition. Hence, the need for magnetic BZs. As we outline in this section, and show in detail in \appref{sec: PerturbationTheory}, the limits of small magnetic field and a flat band conspire to suppress this momentum shift. This allows us to perform perturbation theory to linear order in $B$ and from the perturbed state, we derive the full quantum geometry of the new band.

Our starting point is the Peierls substition, which is known to yield at linear order \cite{ChangNiu96}
\begin{align}\label{HBexpansion}
 \tH
 %
 %
 =H
 +
 e\delta \hat{A}^j v_j+\cO(\delta \A^2)
\end{align}
where $\hbar \vo_j=[-\im\po_j,H]$ and the gauge potential operator is chosen in symmetric gauge $\delta A_j=(\delta B\times \po)_j/2$ with small magnetic field $\delta B_j$. 
Throughout this work, the tilde indicates a quantity perturbed to linear order in $\delta B$. 

Restricting to two dimensions and an out-of-plane magnetic field $\delta B_j= \delta B\delta_{j,z}$, the perturbing operator becomes \cite{ChangNiu96}
\begin{equation}\label{angularMomentum}
 L=\frac{1}{2}e\,\delta B\cdot(\po\times\vo)=\frac{\hbar}{2\LB^2}\varepsilon^{ij}\po_i\vo_j
\end{equation}
with magnetic length $\LB^2=\hbar/|e\delta B|$, which needs to be large compared to the system size. Now, we focus on bands with flat dispersions, $\p_i\epsilon_\alpha=0$. For the remainder of this text, the label $\alpha$ always indicates such a flat band with state $\ket{k\alpha}$; other bands, possibly non-flat, are labelled by $\beta,\gamma$ and so on. Sometimes we refer to $\alpha$ as the \quotes{seed band}.

As shown in detail in \appref{sec: PerturbationTheory} the perturbed energy, state and Bloch function of a flat band are to linear order
\begin{subequations}\label{perturbed}
\begin{align}
 \tepsilon_\alpha(\bfk)&=\epsilon_\alpha+\frac{\m_3^\alpha(\bfk)}{2\LB^2}\label{perturbedEnergy}\\
 \pertket&=\ket{\bfk\alpha}+\frac{1}{2\LB^2}\sum_{\beta\neq\alpha}\ket{\bfk\beta}\,\lcb^{\beta\alpha}(\bfk)\label{perturbedState}\\
 \tu^{\alpha}_a(\bfk)&=u^{\alpha }_a(\bfk)+\frac{1}{2\LB^2}\sum_{\beta\neq\alpha}u^{\beta}_a(\bfk)\lcb^{\beta\alpha}(\bfk)\label{perturbedBloch},
\end{align}
\end{subequations}
We introduced the third component of the orbital magnetic moment already found in semiclassical analyses \cite{ChangNiu96, Hwang},
\begin{equation}
 \m_3^\alpha(\bfk)=\im\sum_{\beta\neq\alpha}(\epsilon_\beta(\bfk)-\epsilon_\alpha)\varepsilon^{ij}\cA^{\alpha\beta}_i(\bfk)\cA^{\beta\alpha}_j(\bfk)\,,
\end{equation}
 and a novel correction
\begin{align}\label{lcb}
 \lcb^{\beta\alpha}(\bfk)&=\varepsilon^{ij}\biggl[
 \frac{(\p_i\epsilon_\beta)(\bfk)}{\epsilon_\beta(\bfk)-\epsilon_\alpha}\cA_j^{\beta\alpha}(\bfk)\notag\\
 &\quad-\im\sum_{\gamma\neq\{\alpha,\beta\}}\left(\frac{\epsilon_\alpha-\epsilon_\gamma(\bfk)}{\epsilon_\alpha-\epsilon_\beta(\bfk)}\right)\cA^{\beta\gamma}_i(\bfk)\cA^{\gamma\alpha}_j(\bfk)
 \biggr]\,.
\end{align}
which controls the mixing of the unperturbed states, when turning on a small magnetic field $\delta B$. 

A number of remarks is in order. Firstly, $\lcb^{\beta\alpha}$ is gauge invariant. Secondly, adopting the convention that the upper indices on $\lcb^{\beta\alpha}$ are primarily associated with the indices on the cross-gap functions, we can write $(\lcb^{\beta\alpha})^*=\lcb^{\alpha\beta}$, i.e. the complex conjugate has identical index placement on the energy ratios. Thirdly, the sum over $\gamma$ only appears for three bands and more. While the first summand describes how the band $\alpha$ couples to a single other band, the second term connects $\alpha$ to two other bands simultaneously. Fourthly, we are still using the momentum $\bfk$ as a quantum number. This owes to the combination of small magnetic field and flat band. The latter suppresses the momentum shifting part of the position operator in \eqref{angularMomentum}. Therefore, the expressions \eqref{perturbedEnergy}-\eqref{perturbedBloch} apply only to flat bands.  

The perturbed Bloch function gives rise to its own Berry connection
\begin{align}\label{ConnectionB}
 \tcA^{\alpha}_j
 =
 \cA^\alpha_j+\LB^{-2}\Gamma_j\,
\end{align}
where
\begin{equation}\label{Gamma}
 \Gamma_j
 =
 \sum_{\beta\neq\alpha}\Re\left[\lcb^{\alpha\beta}\cA^{\beta\alpha}_j\right]
\end{equation}
It is convenient to also define 
\begin{equation}\label{Upsilon}
 \Upsilon_j
 =
 \sum_{\beta\neq\alpha}\Im\left[\lcb^{\alpha\beta}\cA^{\beta\alpha}_j\right]
\end{equation}
Both, $\Gamma$ and $\Upsilon$ are manifestly gauge invariant. Their full expressions with \eqref{lcb} plugged in are presented in \eqref{GammaUpsilonFull} of \appref{sec: PerturbationTheory}. It is then straightforward to derive the Berry curvature of the perturbed band
\begin{subequations}\label{perturbedCurvature}
\begin{align}
 \tOmega^{\alpha}_{ij}(\bfk)
 &=
 2\p_{[i}\tcA_{j]}^{\alpha}(\bfk)
 =
 \Omega_{ij}^\alpha(\bfk)+2\LB^{-2}\p_{[i}\Gamma_{j]}(\bfk)\label{CurvatureBtensor}\\
 \tOmega^{\alpha}(\bfk)
 &=
 \varepsilon^{ij}\tOmega^{\alpha}_{ij}(\bfk)/2
 =
 \Omega^\alpha(\bfk)+\LB^{-2}\varepsilon^{ij}\p_i\Gamma_j(\bfk)\label{CurvatureB}
\end{align}
\end{subequations}
and the quantum metric of the perturbed band
\begin{widetext}
\begin{align}\label{MetricB}
 \tg^{\alpha}_{ij}(\bfk)
%
=g^\alpha_{ij}(\bfk)
+
\LB^{-2}\biggl(
\p_{(i}\Upsilon_{j)}(\bfk)
-
\frac{1}{2}\sum_{\beta\neq\alpha}\biggl[\biggl(
\sum_a(\p_{i}\p_{j} u^{\alpha *}_a) u^{\beta}_a+2\cA^\alpha_{(i}\cA^{\alpha\beta}_{j)}\biggr)\lcb^{\beta\alpha}+c.c.
\biggr]
\biggr)
\end{align}
\end{widetext}
Both, $\tOmega^{\alpha}$ and $\tg^\alpha$, are gauge invariant, the former manifestly so and for the latter this is shown to hold true in \appref{sec: PerturbationTheory}. The gauge invariance of the perturbed quantum geometry is a reassuring feature of our construction. This concludes our derivation of the quantum geometric properties of a flat band perturbed by a small magnetic field.
\section{Ideal Insulators in weak magnetic fields}
\label{sec: IdealInsulatorsB}
Now, we investigate the consequences of the magnetic perturbation in the insulating regime with particular attention on ideal insulators. This presents an application of the connections appearing in \secref{sec: NewConnections}.
We begin in \secref{sec: HallConductivity} by showing that the topological invariant of the system, the Chern number, is unaffected by a small magnetic field. Thereafter, we discuss in turn the three properties of ideal isotropic bands in weak magnetic fields. \Secref{sec: IsolatedFlatBandB} deals with the condition of isolated flatness \eqref{isolatedFlat}, \secref{sec: UniformQGB} with the uniformity of the quantum geometry \eqref{uniformity} and \secref{sec: dropletsB} deals with isotropic droplets \eqref{droplet}. Finally, in \secref{sec: IdealDropletsB} we combine everything and discuss ideal isotropic Chern bands in weak magnetic fields.
\subsection{Hall Conductivity}\label{sec: HallConductivity}
The magnetic perturbation is small, and as such it should leave topological invariants of the system untouched. Indeed, we confirm this expectation right away for the Hall conductivity. 

When computing the Chern number of the perturbed band, it is important to note that $\Gamma$ is gauge invariant while $\cA^\alpha$ is not. After employing Stokes' theorem the contributions to the Chern number from $\cA^\alpha$ result from integrating over the boundary of the $\FBZ$. The $\FBZ$ is a torus and has no boundary. However, $\cA^\alpha$ has poles which need to be removed by small disks, providing a boundary, giving way for a non-vanishing integral; cf. the discussion in \cite{Lee_2017}. The poles in turn result from the gauge \textit{variance} of $\cA^\alpha$. In contrast, repeating the same steps for $\Gamma$, it vanishes as it has no poles by virtue of being gauge invariant. Hence, the Chern number of the perturbed band is the same as that of the seed band,
\begin{equation}\label{ChernNumberB}
 \int_{\FBZ}\d^2k\,\tOmega^{\alpha}(\bfk)
 =
 \int_{\FBZ}\d^2k\,\Omega^{\alpha}(\bfk)
 =
 2\pi C_\alpha.
\end{equation}
An equivalent argument is as follows. The new Berry curvature may be written in form notation as $\tOmega^{\alpha}=\d(\cA^\alpha+\LB^{-2}\Gamma)$. Because $\cA^\alpha$ is gauge variant, the first summand is exact only patchwise. In contrast, because $\Gamma$ is gauge invariant, the form $\d\Gamma$ is globally exact. By Stokes' theorem 
\begin{equation}
 \int_{\FBZ}\d\Gamma=\int_{\p \FBZ}\Gamma=0
\end{equation}
The last equality follows since the $\FBZ$ is a torus, and thus $\p \FBZ=\emptyset$. We stress that this result relies on integration over the full $\FBZ$. In the metallic regime, to be discussed in \secref{sec: Metals}, this property is lifted.
\subsection{Isolated Flat Bands in Weak Magnetic Fields}\label{sec: IsolatedFlatBandB}
The first property of an ideal isotropic Chern band is a that it harbors an isolated flat band, \eqref{isolatedFlat}. It is clear that from \eqref{perturbedEnergy} that the perturbed band is no longer perfectly flat, since
\begin{equation}\label{LandauLevelSpread}
    \p_j\tepsilon_\alpha=(2\LB)^{-2}\p_j\m_3\neq0\,. 
\end{equation}
However, the isolation of the band, $|\epsilon_\beta-\epsilon_\alpha|\gg1$, is not affected, because the magnetic field is taken to be small. The property \eqref{LandauLevelSpread} is called Landau level spread and has been discussed at length in \cite{Hwang}. 

In practice, the flatness only holds in approximation, i.e. there are now two kinds of fluctuations in the band. First, the inherent fluctuations of the band, and second, the fluctuations induced by the Landau level spread \eqref{LandauLevelSpread}. If the Landau level spread outweighs the inherent fluctuations only slightly, the band may still be taken to be a flat isolated band in good approximation.
\subsection{Uniform Quantum Geometry in Weak Magnetic Fields}\label{sec: UniformQGB}
A similar analysis applies for the uniformity of the quantum geometry. Starting from a uniform $\chi^\alpha$ for the seed band, the perturbed curvature \eqref{CurvatureB} and metric \eqref{MetricB} are obviously no longer uniform. In practice, $\chi^\alpha$ fluctuates a little %
\footnote{Note that bands with constant Berry curvature are possible, albeit only for three bands and more \cite{Varjas}. A similar analysis has not been conducted, to our knowledge, for the quantum metric.}. %
These fluctuations have to be compared to those induced by the magnetic field in order to gauge how much the magnetic field affects uniformity.  

A means of quantifying how strongly the uniformity of the quantum geometry is affected by $\delta B$ is provided by the connections in section \secref{sec: NewConnections}. Let us define $\delta_Bg^\alpha=\tg^{\alpha}-g^{\alpha}$ and $\delta_B\Omega=\tOmega^{\alpha}-\Omega^\alpha$, which can be read off from \eqref{MetricB} and \eqref{CurvatureB}, respectively. A perturbed zweibein is introduced in analogy to \eqref{CovariantDerivative}, i.e. 
\begin{equation}
 \tZB_{ja}(\bfk)=(\tD_j\tu_a^\alpha)
\end{equation}
where now the covariant derivative is $\tD_j=\p_j+\im\tcA^\alpha_j$.
Similarly, we define perturbed counterparts $\tU$ and $\tV$ of \eqref{NovelConnections}, which transform as follows under gauge transformations
\begin{align}
 \tU^l&\to \tU^l-\tr(g^{\alpha}+\delta_B g^\alpha)\,\p^l\phi\\
 \tV^l&\to \tV^l+(\Omega^\alpha+\delta_B\Omega^\alpha)\,\p^l\phi
\end{align}
By assumption the unperturbed band is uniform, and thus the second term on the right hand side of each equation can again be absorbed in the function $\phi$. The corrections $\delta_B$, on the other hand, cannot in general be absorbed. Therefore, the new connections provide a means of checking to which degree the magnetic field violates uniformity of the quantum geometry. Since the magnetic field is weak, the uniformity will be violated mildly at most.

In practice, uniformity is present only approximately even for the seed band. As shown in \cite{Parameswaran_2012}, the GMP algebra \eqref{GMP} is realized at long wavelengths $\bfq\ll1$ and $\bfk\ll1$ to quadratic order,
\begin{equation}\label{linearGMP}
 [\rho_{\bfq}^{\alpha},\rho_{\bfk}^{\alpha}]
 =
 \im\,q_i\varepsilon^{ij}k_j\,\corr{\Omega^\alpha}\,
 \rho_{\bfq+\bfk}^{\alpha}
\end{equation}
by employing an averaged Berry curvature
\begin{equation}
 \corr{\Omega^\alpha}=\frac{\int_{\FBZ}\d^2k\Omega^\alpha}{\int_\FBZ \d^2k}=\frac{2\pi C_\alpha}{A_\FBZ}
\end{equation}
where $A_\FBZ$ is the area of the $\FBZ$. The cubic order corrections to the GMP algebra \eqref{linearGMP} demand that the quantum metric also be uniform \cite{Roy_2014}, meaning that $g^\alpha(\bfk)$ is replaced by 
\begin{equation}\label{MetricAverage}
  \corr{g}^\alpha_{ij}=\frac{\int_{\FBZ}\d^2k\,g^\alpha_{ij}}{\int_\FBZ \d^2k}.
\end{equation}
The more uniform the quantum geometry of the band, the better these approximations become. 

Including a small magnetic field $\delta B$, it is clear from \eqref{ChernNumberB} that $\corr{\tOmega^\alpha}=\corr{\Omega^\alpha}$ so that the GMP algebra does not change to quadratic order. However, in general $\corr{\tg}^\alpha_{ij}\neq\corr{g}^\alpha_{ij}$. Hence this presents a deformation of the GMP algebra at cubic order, for ideal Chern bands, i.e. bands satisfying \eqref{isolatedFlat} and \eqref{uniformity}, not however the droplet condition \eqref{droplet}, i.e. ideal isotropic Chern bands. These are discussed in the following. 

\subsection{Isotropic Droplets in Weak Magnetic Fields}\label{sec: dropletsB}
The final and strongest constraint of an ideal Chern band is the isotropic droplet condition \eqref{droplet}. Provided our previous discussion that the weaker requirements of band flatness and quantum geometry uniformity are affected by the magnetic field, it may come as a surprise that the droplet condition is in fact robust against perturbations $\delta B$.
Indeed, as shown in \appref{sec: RobustDroplet}, requiring \eqref{droplet} for the seed band, the perturbed band is also an isotropic droplet,
\begin{align}\label{dropletRobustness}
 \tr\, \tg^{\alpha }=|\tOmega^{\alpha }|
\end{align}
Two remarks are in order for this highly non-trivial result. Firstly, our calculation holds only to linear order in $\delta B$. It is perfectly resonable to expect \eqref{dropletRobustness} to break down once higher order corrections are included.
Secondly, we stress that the robustness \eqref{dropletRobustness} is totally independent of the uniformity constraint \eqref{uniformity}. It does also not require flatness of the perturbed band. However, it relies on the flatness of the seed band, just as all of our perturbation theory.

\subsection{Ideal Isotropic Chern Bands in Weak Magnetic Fields}\label{sec: IdealDropletsB}
So far, we have discussed the the effect of $\delta B$ on the constraints \eqref{isolatedFlat}, \eqref{uniformity} and \eqref{droplet} by themselves. Now we combine all requirements and extract lessons for ideal isotropic Chern bands, in particular geared towards fractional phases of Chern insulators. Once more this is quantified by the band's susceptibility to realize the GMP algebra, which characterizes the incompressibility of the wavefunction, just as in the FQHE. 

Given eq. \eqref{dropletRobustness} it is tempting to think that an ideal isotropic Chern band will satisfy a GMP algebra \eqref{GMP} with $\Omega^\alpha\to\tOmega^\alpha$ and $g^\alpha_{ij}\to\tg^\alpha_{ij}$. It is important to recall, however, that flat seed bands $\alpha$ which are isotropic droplets \eqref{droplet}, cannot realize a uniform quantum geometry exactly \cite{Varjas}, which was the assumption made to reach \eqref{dropletRobustness}. 
Hence, we are forced again to contemplate approximations. The standard choice is to work with the averaged quantum geometry as discussed in \secref{sec: UniformQGB}. In this case, because $\corr{\tOmega^\alpha}=\corr{\Omega^\alpha}$, but $\corr{\tg_{ij}^\alpha}\neq\corr{g_{ij}^\alpha}$ in general. More explicitely, when averaging $\delta_B\Omega$ dies away while $\delta_Bg_{ij}$ does not. Therefore, the result \eqref{dropletRobustness} breaks down once uniformity of the seed band is enforced.
This hinges on the particular choice of approximation made to realize the GMP algebra in the first place, and it stands to reason, whether it is not possible to find better approximation schemes to analyze ideal isotropic Chern insulators.

\section{quantum transport in metals in weak magnetic fields}\label{sec: Metals}
In order to study the effect of the magnetic field in the metallic regime, we compute the semi-classical equations of motion of a wavepacket living on the perturbed band $\pertket$. Thereafter we apply our results to a two-band model.
Here, because we aim to analyze the charge current in the metallic regime, we consider a spatially inhomogeneous electric (however, also a homogeneous field would be enough to see new physics) field as in \cite{Lapa_2019}. 
\begin{equation}
 E^i=E^i_{(0)}+E^{ij}_{(0)}\po_j, \qquad \varphi(\po)=-E^i_{(0)}\po_i-\frac{1}{2}E^{ij}_{(0)}\po_i\po_j
\end{equation}
with $E^{ij}_{(0)}=E^{ji}_{(0)}$. The Hamiltonian is then
\begin{align}\label{HamiltonianBandE}
 \hat{\sfH}=\tH-e\varphi(\po)
\end{align}
A wavepacket on the perturbed band is described by a state 
\begin{equation}\label{wavepacket}
 \ket{\WP}=\sum_{\bfk}\WP(\bfk)\pertket,
\end{equation}
whose amplitude $\WP$ satisfies the normalization condition $\sum_k|\WP(k)|^2=1$. More precisely, it is sharply peaked in momentum space around the momentum carried by the wavepacket's center $\WK$, $|\WP(\bfk)|^2\simeq\delta(k-\WK)$. The locus of the wavepacket's center in space is found by
\begin{align}\label{wavepacketR}
 \WR_j=\bra{\WP}\po_j\ket{\WP}
 \simeq
 \im\frac{(\p_j\WP)(\WK)}{\WP(\WK)}+\tcA^{\alpha}_j(\WK)
\end{align}
The Hamilton function is
\begin{align}\label{wavepacketH}
 \WH&=\bra{\WP}\hat{\WH}\ket{\WP}\notag\\
 &=
 \tepsilon_\alpha(\WK)+eE^i_{(0)}\WR_i+\frac{e}{2}E^{ij}_{(0)}(\WR_i\WR_j+\tg^{\alpha}_{ij}(\WK))
\end{align}
where \eqref{perturbedEnergy} has been employed. Owing to the fact that \eqref{wavepacket}, \eqref{wavepacketR} and \eqref{wavepacketH} all take the same form as in the unperturbed case, the semiclassical equations of motion assume the same form as in \cite{Lapa_2019},
\begin{subequations}\label{eom}
\begin{align}
 \dot{\WR}_i&=\frac{1}{\hbar}\frac{\p\WH}{\p \WK^i}+\frac{e}{\hbar}\tOmega_{ij}^{\alpha }E^j(\WR)\label{eomPosition}\\
 \dot{\WK}_i&=-\frac{e}{\hbar}E_i(\WR)\label{eomMomentum}
\end{align}
\end{subequations}
Note that the particle does not feel a Lorentz force. This is similar to the case of magnetic BZs \cite{ChangNiu96} where the magnetic field is already woven into the band, therefore $\delta B$ appears in the Berry connection.

The electronic current splits into two contributions
\begin{align}
 j_l&=-e\int\frac{\d^2\WK}{(2\pi)^2}f_0(\WK)\dot{\WR}_l=
 j_l^{\text{G}}+j_l^{\text{H}}
 %
%
%
\end{align}
i.e. 
the geometric current \cite{Lapa_2019} and the Hall current, respectively. They are
\begin{align}
 %
 j_l^{\text{G}}&=-\frac{e^2}{2\hbar}\int\frac{\d^2\WK}{(2\pi)^2}f_0(\WK)\left(\p_l\tg^{\alpha}_{ij}\right)(\WK)E^{ij}_{(0)}\label{GeometricCurrent}\\
 j_l^{\text{H}}&=-\frac{e^2}{\hbar}\int\frac{\d^2\WK}{(2\pi)^2}f_0(\WK)\tOmega_{li}^{\alpha }(\WK)E^i(\WR)\label{HallCurrent}
\end{align}
where with $f_0(\WK)$ we restrict to the equilibrium distribution function indicating the occupied states in the BZ at fixed temperature, taken to be zero in this work. 

Formally the geometric and Hall current are the same as their unperturbed counterparts. The novelty is apparent when spelling the perturbed quantum geometry out in terms of the unperturbed quantum geometry via \eqref{CurvatureB} and \eqref{MetricB}. We stress that the corrections are inherently of quantum nature and describe the quantum coupling of the flat band to at most two other bands, see \eqref{lcb}. Hence these corrections are not accessible in the standard semiclassical approach, which contains only couplings to one other band at most. Importantly, the corrections are already present for homogeneous electric fields as visible in \eqref{HallCurrent}; the geometric current vanishes in this case.

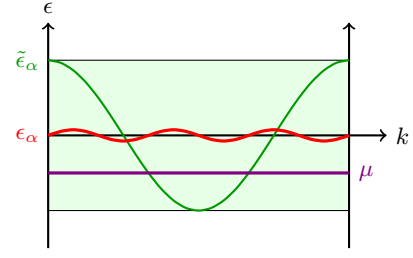
\begin{figure}[t!]
\begin{tikzpicture}
\def\Y{ 1.5}
\def\amp{ 1}
\def\X{ 4}
\def\chem{ -.5*\amp}
 \coordinate (A) at (0,-\amp);
 \coordinate (D) at (0,\amp);
 
 \coordinate (B) at (\X,-\amp);
 \coordinate (C) at (\X,\amp);
  \filldraw[fill=green!10!white, draw=black] (A) -- (B) -- (C) -- (D) -- (A);
  \draw[->, thick] (0, 0) -- (4.5, 0) node[right] {$k$};
  \draw[->, thick] (0, -\Y) -- (0, \Y) node[above] {$\epsilon$};
  \draw[->, thick] (4, -\Y) -- (4, \Y);
  \draw[thick, domain=0:4, smooth, variable=\x, green!60!black] plot ({\x}, {\amp*cos(90*\x)});
  \node[left, green!60!black] at (D) {$\tepsilon_\alpha$};
  \draw[very thick,color=red] plot [domain={0}:{360/90},smooth] (\x,{.075*\amp*sin(270*\x)});
  \node[left, red] at (0,0) {$\epsilon_\alpha$};
  \draw[violet, very thick] (0, \chem) -- (\X, \chem) node[right] {$\mu$};
\end{tikzpicture}
\caption{A flat band $\alpha$ with dispersion $\epsilon_\alpha$ (red) in the $\FBZ$ experiences a LL spread upon applying a weak magnetic field $\delta B$ ending up with dispersion $\tepsilon_\alpha$ (green). Even if the flat band $\epsilon_\alpha$ does not come in contact with a chemical potential $\mu$ (violet), its LL spread may dip or reach into it, rendering the band metallic.}
\label{fig: LLspread}
\end{figure}

If the band is completely filled, the corrections induced by the magnetic field vanish \footnote{In fact, $j^{\text{geom}}=0$ always holds for filled bands.} for the same reason discussed in \secref{sec: HallConductivity}. Hence, such corrections are only be observable in the metallic regime, where the chemical potential $\mu$ lies within the Landau level spread of the perturbed band, $\text{min}(\tepsilon_\alpha)<\mu<\text{max}(\tepsilon_\alpha)$, as sketched in \figref{fig: LLspread}.

\section{A Two-Band Model}\label{sec: Example}
We now discuss how all preceeding expression reduce in the case of two-bands. Thereafter we apply it to an example. The band energy becomes
\begin{align}
 \tepsilon_\alpha(\bfk)
 =
 \epsilon_\alpha+\frac{(\epsilon_\beta(\bfk)-\epsilon_\alpha)}{2\LB^2}\Omega^\alpha(k)
\end{align}
where $\alpha$ is the flat band and $\beta$ the remaining one. This provides the spread of the flat band due to the magnetic field. The corrections \eqref{lcb}, \eqref{Gamma} and \eqref{Upsilon} become
\begin{align}
 \lcb^{\beta\alpha}&=\varepsilon^{ij}\frac{(\p_i\epsilon_\beta)\cA^{\beta\alpha}_j}{\epsilon_\beta-\epsilon_\alpha}\\
 \Gamma_l
 &=
 -g^\alpha_{li}\,\varepsilon^{ij}\frac{\p_j\epsilon_\beta}{\epsilon_\beta-\epsilon_\alpha}\\
 \Upsilon_l
 &=\frac{1}{2}\Omega^{\alpha}
 \frac{\p_l\epsilon_\beta}{\epsilon_\beta-\epsilon_\alpha}
\end{align}
The Berry curvature tensor \eqref{CurvatureB} is corrected by the quantum metric in the two band case,
\begin{align}
 %
 \tOmega^{\alpha }
 =
 \Omega^{\alpha}-\frac{1}{\LB^2}\varepsilon^{ij}\p_{i}\left(g^\alpha_{jk}\varepsilon^{kl}\frac{(\p_l\epsilon_\beta)}{\epsilon_\beta-\epsilon_\alpha}\right)
\end{align}
Similarly, the perturbed quantum metric \eqref{MetricB} is corrected by the Berry curvature and another term in the two band case
\begin{widetext}
\begin{align}\label{MetricBtwoBands}
 \tg_{ij}^{\alpha}
 =
 g_{ij}^{\alpha}
 +
 \frac{1}{2\LB^2}\p_{(i}\left(\Omega^\alpha\frac{\p_{j)}\epsilon_\beta}{\epsilon_\beta-\epsilon_\alpha}\right)
 -
 \frac{1}{2\LB^2}\biggl[\biggl(
\sum_a(\p_{i}\p_{j} u^{\alpha *}_a) u^{\beta}_a+2\cA^\alpha_{(i}\cA^{\alpha\beta}_{j)}\biggr)\lcb^{\beta\alpha}+c.c.\biggr]
\biggr)
\end{align}
These expressions can now be plugged into the electronic currents. We find that the quantum metric of the unperturbed band corrects the Hall current \eqref{HallCurrent}. It is interesting to point out that no non-uniform electric field component $E^{ij}_{(0)}$ is required for the quantum metric to enter. It already appears using a uniform electric field in presence of a uniform magnetic field. The new Hall current is then
\begin{equation}
 j_n^{\text{H}}=-\frac{e^2}{\hbar}\int^\mu\frac{\d^2\WK}{(2\pi)^2}f_0(\WK)
 \left(\Omega^{\alpha}-\frac{1}{\LB^2}\varepsilon^{ij}\p_{i}\left(g^\alpha_{jk}\varepsilon^{kl}\frac{(\p_l\epsilon_\beta)}{\epsilon_\beta-\epsilon_\alpha}\right)\right)
 \varepsilon_{nm}E^m(\WR)
\end{equation}
\end{widetext}
with chemical potential within $\text{min}(\tepsilon_\alpha)<\mu<\text{max}(\tepsilon_\alpha)$. Similarly, plugging \eqref{MetricBtwoBands} into \eqref{GeometricCurrent}, we see that the geometric current is corrected by the Berry curvature, amongst other contributions.

Previously, the quantum metric had only been found to contribute to the longitudinal current as soon as $E^{ij}_{(0)}\neq0$, i.e. the geometric current $j^{\text{G}}$ \cite{Lapa_2019}. Similarly, the Berry curvature corrects the geometric current \eqref{GeometricCurrent}. Clearly, this can only be observed for non-uniform electric field $E^{ij}_{(0)}\neq0$.

\subsection{An Example}
\begin{figure*}[t]
\begin{minipage}{.5\textwidth}
 \includegraphics[scale=.5]{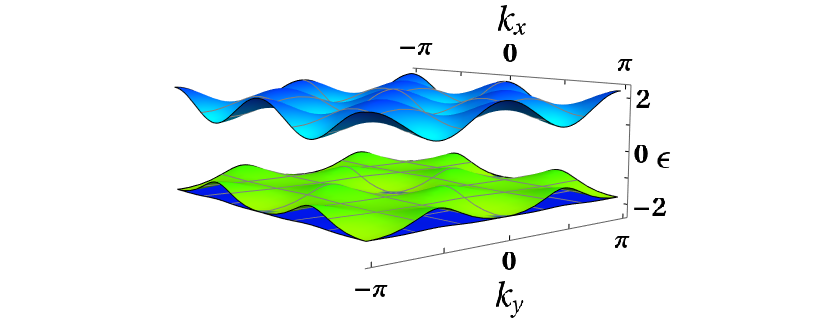}
\end{minipage}
\begin{minipage}{.5\textwidth}
 \includegraphics[scale=.43]{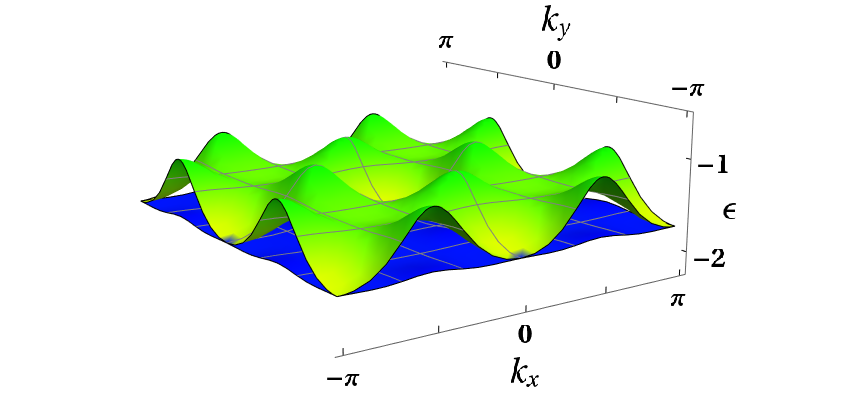}
\end{minipage}
\caption{Left: The bands of the uperturbed system are depicted in blue. The lower band is flat, and thus the candidate for our perturbation theory. Right: Close-up of the unperturbed flat band $\alpha$ in blue and its perturbed counterpart in green. Both plots use flux $\phi/\phi_0=0.1$ and the lattice spacing is set to unity.}
\label{fig: bands}
\end{figure*}

We now exemplify our theoretical concepts using a two-band model $H=d_0\id+\vec{d}\cdot\vec{\sigma}$, where $\vec{\sigma}$ is the vector of Pauli matrices, for a square lattice with a nearly-flat Bloch band introduced in Ref.\cite{Lee_2017},
\begin{align}
 d_0&=0.27\cos(2k_x)\cos(2k_y)\notag\\
 d_1&=\cos(k_x)-\cos(k_y)\notag\\
 d_2&=\cos(k_x)+\cos(k_y)\notag\\
 d_3&=2\sin(k_x)\sin(k_y)
\end{align}
Its lower band is flat giving way for our analysis.
When implementing the magnetic field we use $\LB^{-2}=(2\pi/a_0^2)(\phi/\phi_0)$ where $\phi_0$ is the flux quantum and $\phi=\delta B a_0^2$ the magnetic flux. In this subsection we place the lattice constant $a_0$ at unity and choose a magnetic field with $\phi/\phi_0=0.1$ units of flux. \Figref{fig: bands} displays the two unperturbed bands of $H$ in blue, the lower of which is flat, while the green band is the perturbed flat band with dispersion $\tepsilon_\alpha$. The Landau level spread is clearly visible.

\begin{figure*}[t]
\begin{minipage}{.5\textwidth}
 \raggedleft
 \includegraphics[scale=.45]{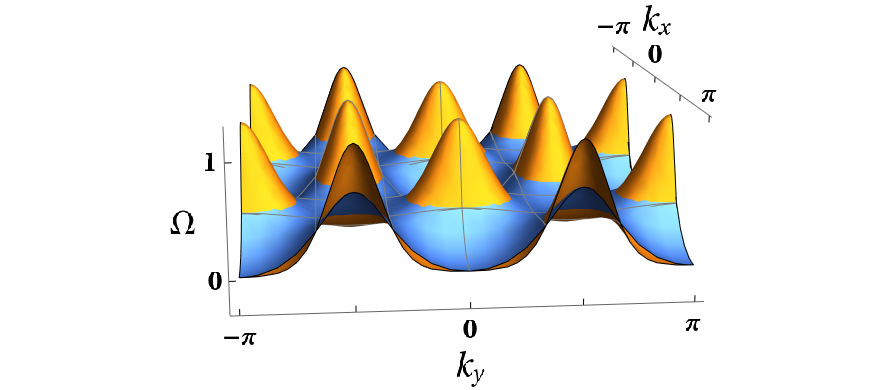}
\end{minipage}
\begin{minipage}{.5\textwidth}
 \begin{center}
 \includegraphics[scale=.4]{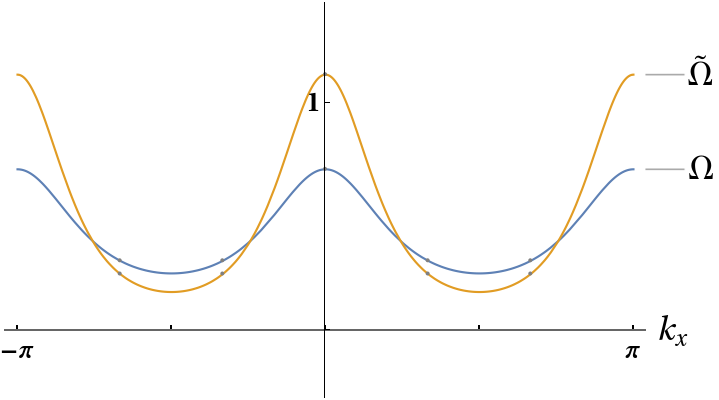}
 \end{center}
\end{minipage}
\caption{Left: The curvatures $\Omega^\alpha$ (blue) and $\tOmega^\alpha$ (orange) are depicted. Right: A section of the metric component in $xx$-direction at fixed $k_y=\pi/2$ is presented. Both plots use flux $\phi/\phi_0=0.1$ and the lattice spacing is set to unity.}
\label{fig: Curvatures}
\end{figure*}

\begin{figure*}[t]
\begin{minipage}{.5\textwidth}
 \raggedleft
 \includegraphics[scale=.45]{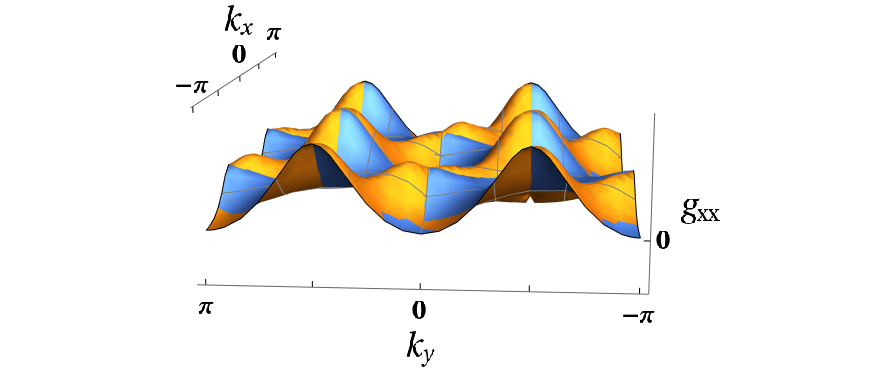}
\end{minipage}
\begin{minipage}{.5\textwidth}
 \begin{center}
 \includegraphics[scale=.4]{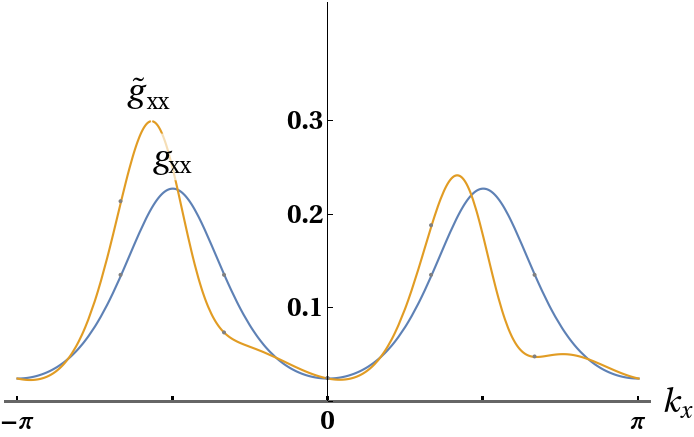}
 \end{center}
\end{minipage}
\caption{Left: The metric component $g^\alpha_{xx}$ (blue) and $\tg^\alpha_{xx}$ (orange) are depicted. Right: A section of the curvatures at fixed $k_y=\pi/2$ is presented. Both plots use flux $\phi/\phi_0=0.1$ and the lattice spacing is set to unity.}
\label{fig: Metrics}
\end{figure*}

As expected from the studies in \cite{Varjas, Lee_2017}, the Berry curvature cannot be uniform for two-band models. We plot both curvatures, once without and with magnetic field in \figref{fig: Curvatures}. The corrections for the quantum metric are very small, as depicted in \figref{fig: Metrics}.

\subsubsection{Emergent Curvatures and Uniformity}
\begin{figure*}[t]
\begin{minipage}{.5\textwidth}
 \raggedleft
 \includegraphics[scale=.35]{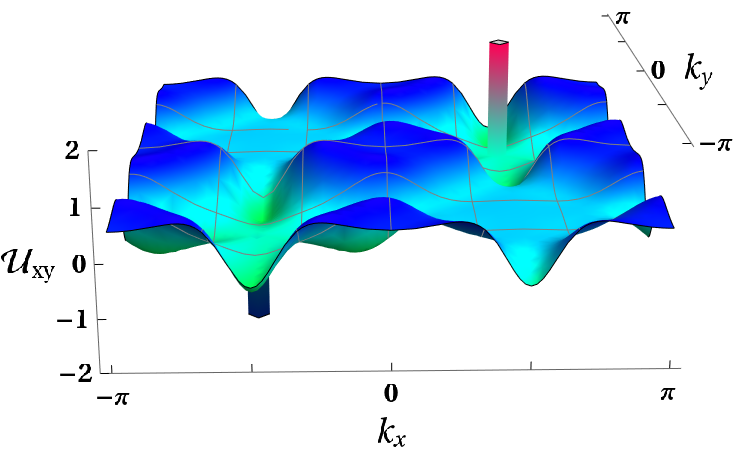}
\end{minipage}
\begin{minipage}{.5\textwidth}
 \begin{center}
 \includegraphics[scale=.4]{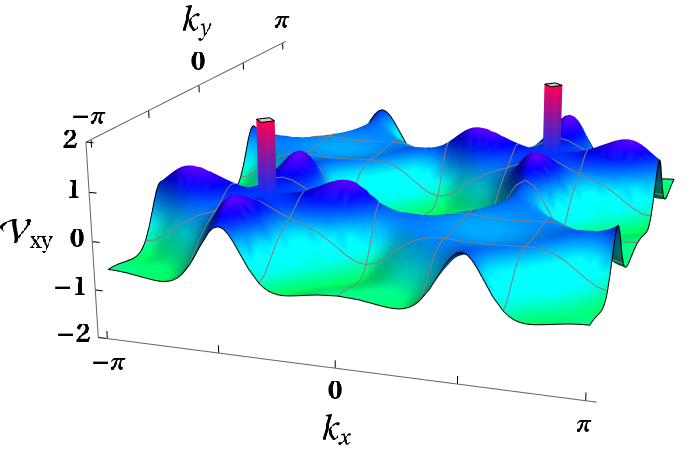}
 \end{center}
\end{minipage}
\caption{The left plot depicts the emergent curvature $\cU_{xy}$ and the right shows $\cV_{xy}$. Both have two poles, corroborating that $\tr(g^\alpha)$ and $\Omega^\alpha$, respectively, are not uniform in the $\FBZ$.}
\label{fig: EmergentCurvatures}
\end{figure*}

The non-uniformity of the quantum geometry is mirrored in the emergent curvatures \eqref{NovelCurvatures}. Both have two poles as depicted in \figref{fig: EmergentCurvatures} indicating that neither $\cU_{xy}$ nor $\cV_{xy}$ are gauge invariant, which is traced back to the fact that $\tr(g^\alpha)$ and $\Omega^\alpha$ vary in the $\FBZ$, respectively. The presence of the magnetic field does not change this circumstance, as expected. Hence $\cV_{xy}$ and $\cU_{xy}$ are not truly curvatures, irrespective of a magnetic field. 

The non-uniformity is quantified by the integrals of \eqref{NovelCurvatures} over the $\FBZ$. Withouth magnetic field they are
\begin{equation}
 2\pi\int_\FBZ \d^2k\,\cU^{xy}=\frac{1}{2}, \quad 2\pi\int_\FBZ \d^2k\,\cV^{xy}=-\frac{1}{2}\,.
\end{equation}
and with magnetic field
\begin{equation}
 2\pi\int_\FBZ \d^2k\,\tcU^{xy}=0.35, \quad 2\pi\int_\FBZ \d^2k\,\tcV^{xy}=-0.35\,.
\end{equation}
Neither integral is an integer confirming the non-uniformity of the quantum geometry.
It is interesting that with and without magnetic field we obtain 
\begin{equation}
 2\pi\int_\FBZ \d^2k(\cV^{xy}+\cU^{xy})=0\,.
\end{equation}
For now we leave it as an observation to be investigated in the future.

\section{Discussion}\label{sec: conclusions}
\subsection{Conclusions}
Summarizing, in this work we have analyzed novel aspects of ideal Chern insulators in two spatial dimensions. Firstly, we have shown that new effective Berry connections can be consistently defined in these systems once the band's quantum geometry, namely the Berry curvature and the quantum metric, is uniform.
In this case, quantized Chern numbers arise from the curvatures of our novel connections. Hence, these connections provide a measure for the uniformity of the quantum geometry. 

Secondly, we studied nearly-flat band systems, not necessarily ideal, in the presence of electromagnetic fields. We provided deformations of the Berry curvature and quantum metric upon applying a small magnetic field and went on investigating the properties of this new quantum geometry in the insulating and metallic regime. We point out that our work is to be distinguished from recent advances implementing strong magnetic fields which push the system into the Hofstadter phase \cite{Scaffidi, bauer2021fractional}.

In the insulating regime, we showed that the Hall conductivity is not affected by small magnetic fields, in agreement with previous work. Our next steps were to investigate the influence of this weak magnetic field on the three ingredients of ideal isotropic Chern bands. Given an ideal isotropic Chern band, upon applying a small magnetic field, the flatness and the uniformity of the quantum geometry are affected. The former has been investigated in detail in \cite{Hwang}, since the energy correction we found is identical with that of the semiclassical approach in \cite{ChangNiu96}. The deformation of the quantum geometry away from uniformity provides an application of the emergent connections discussed before. They indicate how strongly the uniformity is affected by the weak magnetic field. Next, we demonstrated the highly non-trivial fact that ideal isotropic droplets are robust against perturbation with a magnetic field in the following sense: Starting from a quantum geometry satisfying the droplet condition \eqref{droplet}, the perturbed quantum geometry satisfies \eqref{dropletRobustness}. Putting everything together, we discussed how ideal isotropic droplets are affected by magnetic fields. 

Moving to the metallic regime and turning on an electric field, we determined the modifications caused by the weak magnetic field to the semiclassical equations governing electronic motion on the band. We point out that, even though a magnetic field is turned on, the electron does not experience a Lorentz force, at least not directly. The situation is similar to the Hofstadter case, where the magnetic field is woven into the band's Berry curvature \cite{ChangNiu96}, and more generally its quantum geometry. Therefore the electron experiences the effect of the magnetic field through the inherent properties of the band itself, not as an external force. Furthermore, we derived the corrections to the electronic current, which is split into a longitudinal geometric \cite{Lapa_2019} and transversal Hall current. In the two-band case, we showed how the Hall current is corrected by the quantum metric and the geometric current is corrected, amongst other terms, by the Berry curvature of the seed flat band. More generally, for any number of bands, the corrections are determined by the so-called quantum geometric interband fidelities, see \eqref{fidelities} and \eqref{GammaUpsilonFull} of \appref{sec: PerturbationTheory}.

\subsection{Outlook}
Ideal Chern insulators have first been discussed in the context of fractional Chern insulators \cite{RegnaultBernevig, Neupert_2015, ThomaleDuality}, which aim to mimick the FQHE in lattice systems. Our work dealt exclusively with properties of ideal bands and it is an interesting future endeavor to study the consequences of our work once interactions are implemented on the band. This is similar in spirit, yet significantly different from \cite{bauer2021fractional}, which presents as well a study of fractional Chern insulators in magnetic fields. Our work differs from theirs, since they employed strong magnetic fields $B$ allowing them to use magnetic Brillouin zones and work entirely with its quantum geometry. This is the analog of our effective quantum geometry. Their approach is viable only in the case of rational, yet large, magnetic flux. In contrast our work is viable for non-rational flux as well, albeit for weak magnetic flux. It will be interesting to combine their work with ours, similar to \cite{ChangNiu96}, which employed a strong magnetic field $B$ like \cite{bauer2021fractional} to reach the Hofstadter phase and perturbed additionally with a small magnetic field $\delta B$ like ours and of non-rational flux. Taken together the magnetic field $B+\delta B$ should give rise to a new quantum geometry governing the properties of the interacting system subject to large non-rational magnetic flux. Another avenue is to include a confining potential to induce momentum space duals of LLs \cite{ThomaleDuality}. This paper furthermore showed how interactions in ideal isotropic bands are mapped into Haldane's geometric description \cite{Haldane_2011} and it will be interesting to investigate how small magnetic fields affect their analysis. We are convinced that our results will be central in the correct understanding of fractional Chern insulating phases in presence of external electromagnetic perturbations as we will show in future work. 


A very intriguing experimental platform to test our results is twisted bilayer graphene \cite{Xie, Rossi, Julku, Wu, AhnSuperconductivity, Ledwith, Toermae, Hu}. It demands the improvement of our theory to degenerate flat bands. Therefore the quantum geometry results from non-abelian gauge theory. Furthermore, we point out that for twisted bilayer graphene there is an interesting connection between the GMP algebra of an ideal flat band and the superfluid weight. Indeed, the average of the metric \eqref{MetricAverage}, governing the GMP algebra at cubic order, is proportional to the superfluid weight \cite{Xie}.

Irrespective of ideal bands, our perturbation theory is based on the flatness of the band, so that our findings can be applied to realistic flat band models, for instance rhombohedral trilayer graphene at low energies. It harbors six bands, two of which are almost flat within a momentum range of $|\bfk|\leq0.05a_0$, where $a_0$ is the lattice constant \cite{Mitscherling}. In this case, our formalism can be applied to both flat bands so that we can in fact determine the effective quantum geometry of more than a single band. More importantly it provides a measurable avenue for our results. Going beyond our work, it is interesting to see how susceptible our perturbation theory is to response techniques such as those of \cite{Mitscherling}. We suspect that, just as in \eqref{eom}, it will be legitimate to employ our effective quantum geometry instead of the seed band's quantum geometry, providing immediate corrections to their response functions.

Before closing, we mention a few different directions. Firstly, it is interesting to investigate how the quantum geometry is affected by weak magnetic fields in dimensions higher than than two. This is beneficiary since analyses such as the one in \cite{PriceZilberberg} iterate the linear order semiclassical equations of motion to reach quadratic order in electromagnetic fields. This may be repeated with analogs of our quantum mechanically equations of motion \eqref{eom}. Secondly, it is also intriguing to contemplate lower dimensions such as in \cite{Li}. Thirdly, it is interesting to relate our formalism to that in \cite{Raoux}, which is based on perturbation theory of Green's functions for small magnetic fields. Finally, it would be interesting to study the effect of weak magnetic fields on the GMP algebra in non-hermitian Chern insulators.


\begin{acknowledgments}
It is our pleasure to thank Raffael Klees, Ronny Thomale and Chunxu Zhang for helpful correspondence. E.M.H. and C.N. acknowledge support by the Deutsche Forschungsgemeinschaft (DFG, German Research Foundation) under Germany's Excellence Strategy through the W\"urzburg‐Dresden Cluster of Excellence on Complexity and Topology in Quantum Matter ‐ ct.qmat (EXC 2147, project‐id 390858490). The work of E.M.H. and C.N. was furthermore supported via project id 258499086 - SFB 1170 ’ToCoTronics’.
\end{acknowledgments}
\bibliographystyle{apsrev4-1}
\bibliography{Draft} 

\newpage
\begin{widetext}
\appendix 
\section{Perturbation Theory}\label{sec: PerturbationTheory}
In the main text, we present results for the perturbation theory of the operator $L=\hbar\LB^{-2}\varepsilon^{ij}\po_i\vo_j/2$ about the unperturbed Hamiltonian $H$. This requires knowledge of the matrix element \eqref{POband} of the main text and also of the velocity operator $\vo_j=\im[H,\po_\mu]$,
\begin{align}\label{VObandBasis}
 \bra{\bfq\beta}\hbar \vo_j\ket{\bfk\alpha}
 =\delta(\bfq-\bfk)\biggl(
 \delta^{\alpha\beta}(\p_j\epsilon_\alpha)(\bfk)+\im(\epsilon_\beta(\bfk)-\epsilon_\alpha(\bfk))\cA^{\beta\alpha}_j(\bfk)
 \biggr)
\end{align}
The matrix element of the angular momentum operator can then be evaluated
\begin{align}
\bra{\bfq\beta}L\ket{\bfk\alpha}&=\frac{\hbar}{2\LB^2}\varepsilon^{ij}\sum_{\bfp,\gamma}\bra{\bfq\beta}\po_i\ketbra{\bfp\gamma}{\bfp\gamma}\vo_j\ket{\bfk\alpha}\notag\\
&=\frac{1}{2\LB^2}\varepsilon^{ij}\biggl[-
\frac{\p}{\p k^i}(\delta(\bfq-\bfk))\biggl(
 \im\delta^{\beta\alpha}(\p_j\epsilon_\alpha(\bfk))-(\epsilon_\beta(\bfk)-\epsilon_\alpha(\bfk))\cA^{\beta\alpha}_j(\bfk)\biggr)\notag\\
 &\qquad+
 \delta(\bfq-\bfk)\biggl(\cA_i^{\beta\alpha}(\bfk)(\p_j\epsilon_\alpha(\bfk))+\im\sum_\gamma(\epsilon_\gamma(\bfk)-\epsilon_\alpha(\bfk))\cA^{\beta\gamma}_i\cA^{\gamma\alpha}_j(\bfk)\biggr)
\biggr]\label{LinearOrderPerturbation}\\
&\to\frac{1}{2\LB^2}\varepsilon^{ij}\biggl[
\frac{\p}{\p k^i}(\delta(\bfq-\bfk))(\epsilon_\beta(\bfk)-\epsilon_\alpha(\bfk))\cA^{\beta\alpha}_j(\bfk)\notag\\
 &\qquad+
 \delta(\bfq-\bfk)\im\sum_\gamma(\epsilon_\gamma(\bfk)-\epsilon_\alpha(\bfk))\cA^{\beta\gamma}_i\cA^{\gamma\alpha}_j(\bfk)\label{LinearOrderPerturbationFlat}
\biggr]
\end{align}
where in the last line a flat band, $\p_j\epsilon_\alpha=0$, is assumed. The energy correction to the $\alpha^\th$ band at first order in perturbation theory is $\bra{\bfk\alpha}L\ket{\bfk\alpha}$. For future reference, it is useful to consider instead
\begin{equation}\label{linearEnergyCorrection}
 \bra{\bfq\alpha}L\ket{\bfk\alpha}
=
\frac{1}{2\LB^2}\delta(\bfq-\bfk)
 \,\im\sum_{\gamma\neq\alpha}(\epsilon_\gamma(\bfk)-\epsilon_\alpha(\bfk))\varepsilon^{ij}\cA^{\alpha\gamma}_i\cA^{\gamma\alpha}_j(\bfk)
 =
 \delta(\bfq-\bfk)\frac{\m_3^\alpha(\bfk)}{2\LB^2}
\end{equation}
This matrix element is diagonal in momentum space.
If it were not for the flat band limit, a momentum shifting term still present in the first line of \eqref{LinearOrderPerturbation} would force a shift in momentum, as expected from the Peierls substition, and prevent the evaluation of the matrix element. Its absence lets us proceed with $\bfk$ as a good quantum number here. Hence, the energy correction $\bra{\bfk\alpha}L\ket{\bfk\alpha}$ presented in the main text in \eqref{perturbedEnergy} is obtained simply by dropping the delta distribution. 

The state $\ket{\bfk\alpha}$ is corrected by
\begin{align}
 \sum_{\ket{\bfq\beta}\neq\ket{\bfk\alpha}}\frac{\bra{\bfq\beta}L\ket{\bfk\alpha}}{\epsilon_\alpha(\bfk)-\epsilon_\beta(\bfq)}\ket{\bfq\beta}
 =\sum_{\bfq,\beta\neq\alpha}\frac{\bra{\bfq\beta}L\ket{\bfk\alpha}}{\epsilon_\alpha(k)-\epsilon_\beta(q)}\ket{q\beta}
 +
 \sum_{q\neq k} \frac{\bra{q\alpha}V\ket{k\alpha}}{\epsilon_\alpha(k)-\epsilon_\beta(q)}\ket{q\beta}\biggl|_{\beta\to\alpha}
\end{align}
Since we chose the band to be flat the second term appears to be divergent and needs to be treated with degenerate perturbation theory. The standard protocoll is to find a basis in which the matrix elements $\bra{q'\alpha}V\ket{k'\alpha}$, vanish for $\ket{q'\alpha}\neq\ket{k'\alpha}$. This is the case if the matrix $\bra{q'\alpha}V\ket{k'\alpha}$ is diagonal in momentum space. As we have seen in \eqref{linearEnergyCorrection} this is already true here by the virtue of the flat band approximation. Thus we can drop the second term. The first term can be massaged further by plugging in \eqref{LinearOrderPerturbationFlat} and evaluating the sum over $q$,
\begin{align}\label{lcbAppendix}
 \frac{1}{2\LB^2}\varepsilon^{\mu\nu}\sum_{\beta\neq\alpha}\biggl[
 \frac{(\p_\mu\epsilon_\beta)(k)}{\epsilon_\beta(k)-\epsilon_\alpha}\cA_\nu^{\beta\alpha}(k)
 -\im\sum_{\gamma\neq\{\alpha,\beta\}}\left(\frac{\epsilon_\alpha-\epsilon_\gamma(k)}{\epsilon_\alpha-\epsilon_\beta(k)}\right)\cA^{\beta\gamma}_\mu\cA^{\gamma\alpha}_\nu
 \biggr]\ket{k\beta}
 \equiv \frac{1}{2\LB^2}\sum_{\beta\neq\alpha}\ket{k\beta}\,\lcb^{\beta\alpha}(k)\,.
\end{align}
This requires 
\begin{equation}\label{Subtle}
 \underbrace{\sum_\bfq\ketbra{\bfq\beta}{\bfq\beta}}_{\id_{\beta}}(\p_j\ket{\bfk\beta})=-\im\cA^{\beta}_j(\bfk)\ket{\bfk\beta}
\end{equation}
where $\id_\beta$ is the unit operator on the $\beta^\th$ band and periodicity of the Bloch functions is used to yield the Berry connection and a delta distribution, which eliminates the sum. Note that the contribution \eqref{Subtle} is not gauge invariant and is in fact needed to cancel the gauge variance in the remaining terms, thus leading to the gauge invariance of $\lcb^{\beta\alpha}$ used extensively in the main text. This provides \eqref{perturbedState} and  \eqref{perturbedBloch} of the main text.

It is useful to define the quantum geometric interband fidelities functions for $\beta\neq\alpha$
\begin{subequations}\label{fidelities}
\begin{align}
 \chi_{ij}^{\alpha\beta}&=\cA^{\alpha\beta}_i\cA^{\beta\alpha}_j\\
 g^{\alpha\beta}_{ij}&=\Re\bigl(\cA^{\alpha\beta}_i\cA^{\beta\alpha}_j\bigr)\\
 \Omega^{\alpha\beta}_{ij}&=-2\Im\bigl(\cA^{\alpha\beta}_i\cA^{\beta\alpha}_j\bigr)
\end{align}
\end{subequations}
Recalling \eqref{QGband}, it is clear that the quantum geometric interband fidelities specify how much each band $\beta\neq\alpha$ contributes to the quantum geometry of $\alpha$. In the case of two bands only, the interband fidelities are already the quantum geometry.

This allows to provide a formula for \eqref{Gamma} and \eqref{Upsilon} of the main text, respectively,
\begin{subequations}\label{GammaUpsilonFull}
\begin{align}
 \Gamma_i
 &=
 \sum_{\beta\neq\alpha}\Re\left[\lcb^{\alpha\beta}\cA^{\beta\alpha}_i\right]
 =
 \sum_{\beta\neq\alpha}\varepsilon^{jl}\biggl[
 \frac{\p_j\epsilon_\beta}{\epsilon_\beta-\epsilon_\alpha}g^{\alpha\beta}_{il}
 -\sum_{\gamma\neq\{\alpha,\beta\}}\frac{\epsilon_\alpha-\epsilon_\gamma}{\epsilon_\alpha-\epsilon_\beta}\Im(\cA^{\alpha\beta}_i\cA_l^{\beta\gamma}\cA^{\gamma\alpha}_j)
 \biggr]\\
 \Upsilon_i
 &=
 \sum_{\beta\neq\alpha}\Im\left[\lcb^{\alpha\beta}\cA^{\beta\alpha}_i\right]
 =
 \sum_{\beta\neq\alpha}\varepsilon^{jl}\biggl[
 \frac{\p_j\epsilon_\beta}{\epsilon_\beta-\epsilon_\alpha}\frac{\Omega^{\alpha\beta}_{il}}{2}
 -\sum_{\gamma\neq\{\alpha,\beta\}}\frac{\epsilon_\alpha-\epsilon_\gamma}{\epsilon_\alpha-\epsilon_\beta}\Re(\cA^{\alpha\beta}_i\cA_l^{\beta\gamma}\cA^{\gamma\alpha}_j)
 \biggr]
\end{align}
\end{subequations}
These equations confirm the claim in the main text that the perturbations draw on couplings with at most two other bands. The first summand in $\Gamma$ and $\Upsilon$ describes the coupling with one other band $\beta$, while the second summand accessess two other bands $\beta, \,\gamma$.
To conclude this appendix, we demonstrate the gauge invariance of \eqref{MetricB}. Since $g^\alpha$, $\Gamma$ and $\lcb^{\beta\alpha}$ are gauge invariant by themselves, it is only necessary to inspect
\begin{equation}
\sum_a(\p_{i}\p_{j} u^{\alpha *}_a) u^{\beta}_a+2\cA^\alpha_{(i}\cA^{\alpha\beta}_{j)}, \qquad \beta\neq\alpha
\end{equation}
Indeed, under a gauge transformation $u^\gamma_a\to \e^{\im\phi}u^\gamma_a$, where $\gamma=\alpha,\beta$, the individual summands transform as
\begin{equation}
 \sum_a(\p_{i}\p_{j} u^{\alpha *}_a) u^{\beta}_a\to\sum_a(\p_{i}\p_{j} u^{\alpha *}_a) u^{\beta}_a+2(\p_{(i}\phi)\cA^{\beta\alpha}_{j)}, 
 \qquad
 2\cA^\alpha_{(i}\cA^{\alpha\beta}_{j)}\to 2\cA^\alpha_{(i}\cA^{\alpha\beta}_{j)}-2(\p_{(i}\phi)\cA^{\beta\alpha}_{j)}
\end{equation}
where the orthonormality of the Bloch wave functions $\sum_au^{\alpha *}_au^\beta_a=0$ for $\beta\neq\alpha$ was used.
\section{Robustness of Droplets}\label{sec: RobustDroplet}
In this section the robustness of the droplet condition against magnetic perturbation, equation \eqref{dropletRobustness}, is derived. We proceed in two steps. The first step, found in \appref{sec: IdealCrossGap}, is to show that the ideal droplet condition is equivalent to the vanishing of cross gap functions in complex coordinates. Subsequently, as presented in \appref{sec: RobustDropletProof}, this condition allows to proof the droplet condition. Our conventions on complex coordinates are found in \appref{sec: complexConventions}.
\subsection{Vanishing Cross-Gap Functions in Ideal Isotropic Bands}\label{sec: IdealCrossGap}
The power of the condition \eqref{droplet} is best appreciated in complex coordinates, where it is equivalent to
\begin{align}\label{dropletCrossGap}
 \cA^{\beta\alpha}_z(\bfk)=0,\quad & \cA^{\alpha\beta}_{\bz}(\bfk)=0, \text{ for } \beta\neq\alpha \text{ and }\Omega^\alpha(\bfk)>0,\\
 \cA^{\beta\alpha}_{\bz}(\bfk)=0,\quad & \cA^{\alpha\beta}_{z}(\bfk)=0, \text{ for } \beta\neq\alpha \text{ and }\Omega^\alpha(\bfk)<0,\notag
\end{align}
It is the purpose of this appendix to derive this fact.

The starting point is a result of \cite{Roy_2014},
\begin{align}
 \bra{\bfq\alpha}(Q^\alpha\po_{z}P^\alpha)^\dagger Q^\alpha \po_{z}P^\alpha\ket{\bfk\alpha}\geq0, \text{  if  } \Omega^\alpha>0\label{idealGQcondPosBerry}\\
 \bra{\bfq\alpha}(Q^\alpha\po_{\bz}P^\alpha)^\dagger Q^\alpha \po_{\bz}P^\alpha\ket{\bfk\alpha}\geq0, \text{  if  } \Omega^\alpha<0\label{idealGQcondNegBerry}
\end{align}
where $P^\alpha=\sum_\bfk\ketbra{\bfk\alpha}{\bfk\alpha}$ and $Q^\alpha=\id-P^\alpha$ are band projectors. We discuss here only the case of positive Berry curvature; the case of negative Berry curvature is found by swapping $(z\leftrightarrow\bz)$. If \eqref{droplet} holds, then \eqref{idealGQcondPosBerry} saturates. In other words, for an arbitrary state of the system,
\begin{equation}\label{testKet}
 \ket{\psi}=\sum_{\bfk,\beta}\psi_\beta(\bfk)\ket{\bfk\beta}
\end{equation}
the norm 
\begin{equation}
 || Q^\alpha\po_z P^\alpha\ket{\psi} ||^2=\sum_{\bfk\bfq}\psi_\alpha^*(q)\bra{\bfq\alpha}(Q^\alpha\po_{z}P^\alpha)^\dagger Q^\alpha \po_{z}P^\alpha\ket{\bfk\alpha}\psi_\alpha(\bfk)=0
\end{equation}
vanishes. Note that $P^\alpha$ eleminates the sum over bands. Because $\ket{\psi}$ is arbitrary, operator identities are extracted
\begin{equation}\label{dropletOperator}
 Q^\alpha\po_z P^\alpha=0,\qquad P^\alpha\po_{\bz} Q^\alpha=0\,.
\end{equation}
This constraint guarantees that the operator $\po_z$ does not leak out of the $\alpha^\th$ band and that $\po_{\bz}$ does not leak into it. It is equivalent to
\begin{equation}\label{dropletOperatorSum}
 \sum_{\beta\neq\alpha}P^\beta\po_zP^\alpha=0
\end{equation}
Taken at face value, it means that the contribution over all bands $\beta\neq\alpha$ are non-vanishing but cancel out. In fact, it is possible to show that $\po_z$ does not mix any bands, i.e. the contributions in the sum \eqref{dropletOperatorSum} vanish on their own. Indeed, consider a state $\ket{\phi}$ of type \eqref{testKet}, project it to a band $\beta\neq\alpha$ and sandwhich the operator identity \eqref{dropletOperator} between this state and another test ket $\ket{\psi}$
\begin{equation}
 0=(\bra{\phi}P_\beta)Q^\alpha\po_zP^\alpha\ket{\psi}=\sum_{\bfk}\phi_\beta^*(\bfk)\cA^{\beta\alpha}_z(\bfk)\psi_\alpha(\bfk)
\end{equation}
where \eqref{dropletOperatorSum} and \eqref{POband} have been employed. Hence, the bands $\alpha$ and $\beta$ cannot communicate via $\po_z$. Since the states $\ket{\phi}$ and $\ket{\psi}$ are arbitrary, they can be stripped off and what remains is
\begin{equation}\label{dropletCrossGapAppendix}
 \cA^{\beta\alpha}_z=0, \qquad \cA^{\alpha\beta}_{\bz}=0, \qquad \beta\neq\alpha
\end{equation}
as promised. Hence as seen from \eqref{POband}, the position operator $\po_z$ never leads out of the band $\alpha$, just as the ladder operators incrementing angular momentum inside a LL in the QHE. Taking once with $\p_z$ and once with $\p_{\bz}$ on \eqref{dropletCrossGapAppendix} gives the following useful relation for isotropic droplet Bloch functions
\begin{align}
 \sum_a(\p_z\p_{\bz}u^{\alpha *}_a)u^\beta_a&=-\sum_a(\p_{\bz}u^{\alpha *}_a)(\p_zu^\beta_a),\label{derivativeCrossGap}\\
 \sum_au^{\beta *}_a(\p_z\p_{\bz}u^{\alpha}_a)&=-\sum_a(\p_{\bz}u^{\beta *}_a)(\p_zu^\alpha_a)
\end{align}

Furthermore, a glance at \eqref{QGband} reveals that in this case the quantum geometry for band $\alpha$ has only a single non-vanishing component, the others serving as constraint determining the final component in complex coordinates,
\begin{equation}\label{dropletComplexQG}
 \chi_{z\bz}^\alpha(\bfk)=2g_{z\bz}^\alpha(\bfk)=-\im\Omega_{z\bz}^\alpha(\bfk)=\Omega^\alpha(\bfk)/2
\end{equation}
The fact that the metric is off-diagonal in complex coordinates and thus diagonal in Cartesian coordinates justifies the epiphet \quotes{isotropic} for the droplet condition. Finally, \eqref{Gamma} and \eqref{Upsilon} reduce in a droplet to
\begin{align}
 \Upsilon_z&=\frac{\im}{2}\sum_{\beta\neq\alpha}\cA^{\alpha\beta}_z\lcb^{\beta\alpha}=\im\Gamma_z\\
 \Upsilon_{\bz}&=-\frac{\im}{2}\sum_{\beta\neq\alpha}\lcb^{\alpha\beta}\cA^{\beta\alpha}_{\bz}=-\im\Gamma_{\bz}
\end{align}
\subsection{Proof of Robustness of Droplets}\label{sec: RobustDropletProof}
At last, we are in shape to proof the robustness of the isotropic droplet condition \eqref{dropletRobustness} against small magnetic perturbations. In our conventions we have $S_j T^j=2(S_{z}T_{\bz}+S_{\bz}T_z)$ and $S_j S^j=4S_zS_{\bz}$ for tensors $S,\,T$. Taking a look at \eqref{LeviCivita} it is simple to see that
\begin{equation}\label{partialGammaUpsilon}
 \p_j\Upsilon^j=2\im(-\p_{z}\Gamma_{\bz}+\p_{\bz}\Gamma_z)=\varepsilon^{ij}\p_{i}\Gamma_j
\end{equation}
and $\p_j\p^j=4\p_z\p_{\bz}$. Tracing \eqref{MetricB} gives 
\begin{equation}\label{PreCalc}
 \tg^{\alpha}{}_i{}^{i}
 =
 g^{\alpha }{}_i{}^{i}+\LB^{-2}\varepsilon^{ij}\p_{i}\Gamma_j
 +
 \frac{2}{\LB^2}\sum_{\beta\neq\alpha}\biggl[\biggl(\sum_a(\p_{\bz}u^{\alpha *}_a)(\p_{z}u^{\beta}_a)-\cA_z^\alpha\cA^{\alpha\beta}_{\bz}\biggr)\lcb^{\beta\alpha}+\text{c.c.}\biggr]
\end{equation}
where \eqref{derivativeCrossGap} and \eqref{partialGammaUpsilon} have been employed. The first two summands already give \eqref{CurvatureB}. By plugging in \eqref{CrossGap} it is easy to see that
\begin{equation}
 \sum_a(\p_{\bz}u^{\alpha *}_a)(\p_{z}u^{\beta}_a)-\cA_z^\alpha\cA^{\alpha\beta}_{\bz}
 =
 \sum_{ab}(\p_{\bz}u^{\alpha *}_a)(\delta_{ab}-u^\alpha_au^{\alpha *}_b)(\p_{z}u^{\beta}_b)
 =
 \frac{2}{\LB^2}\sum_{\beta\neq\alpha}\sum_{\gamma\neq\alpha}\cA^{\alpha\gamma}_{\bz}\cA^{\gamma\beta}_z\lcb^{\beta\alpha}
 =
 0
\end{equation}
where the projector $Q^\alpha_{ab}=\sum_{\gamma\neq\alpha}u^{\gamma}_au^{\gamma *}_b$ was used. The last line is reached by employing the isotropic droplet condition \eqref{dropletCrossGap} once more.
Hence, we arrive at our claim, equation \eqref{dropletRobustness} in the main text, stated here for convenience
\begin{align}
 \tr \tg^{\alpha }
 %
 %
 &=\tOmega^{\alpha}
\end{align}
This derivation works similarly for $\Omega^\alpha<0$ with $z$ and $\bz$ interchanged. This means that a flat isotropic droplet remains an isotropic droplet even after perturbing with a small magnetic field, as claimed in the main text. 
\section{Some Conventions on Complex Coordinates}\label{sec: complexConventions}
For the readers convenience, we collect here some conventions on our choice of complex structure. The complex coordinate operators and their conjugate are
\begin{align}
 \po^z&=\hat{z}=\hat{x}+\im\hat{y}=\po^x+\im\po^y,\\ 
 \po^{\bz}&=\hat{\bz}=\hat{x}-\im\hat{y}=\po^x-\im\po^y
\end{align}
While in cartesian coordinates the metric is $\eta_{ij}=\text{diag}(1,1)$, and the Levi-Civita symbol is $\varepsilon_{xy}=-\varepsilon_{yx}=1$, in complex coordinates they are
\begin{equation}\label{LeviCivita}
 \eta_{ij}=\begin{pmatrix}
                0 & \frac{1}{2}\\
                \frac{1}{2} & 0
               \end{pmatrix}
\qquad
\eta^{ij}=\begin{pmatrix}
                0 & 2\\
                2 & 0
               \end{pmatrix}
\qquad
\varepsilon_{ij}=\begin{pmatrix}
                0 & \frac{i}{2}\\
                -\frac{i}{2} & 0
               \end{pmatrix}
\qquad
\varepsilon^{ij}=\begin{pmatrix}
                0 & -2i\\
                2i & 0
               \end{pmatrix}
\end{equation}
Since our interest lies on 1-forms, we lower an index $\po_i=\eta_{ij}\po^j$, i.e. $\po_z=\hat{\bz}/2=\po^{\bz}/2$ and $\po_{\bz}=\hat{z}/2=\po^{z}/2$. 
The complex components of 1-forms, we collectively call them $\cA$, having cross-gap functions and Berry connections in mind, are then
\begin{align}
 \cA_z=\frac{1}{2}(\cA_x-\im\cA_y) \quad \cA_{\bz}=\frac{1}{2}(\cA_x+\im\cA_y)
\end{align}
Clearly, $(\cA^{\beta\alpha}_{\bz})^*=\cA^{\alpha\beta}_z,\, (\cA^{\beta\alpha}_{z})^*=\cA^{\alpha\beta}_{\bz}$.

\end{widetext}

\end{document}